\documentclass[conference]{IEEEtran}
\IEEEoverridecommandlockouts
\usepackage{cite}
\usepackage{amsmath,amssymb,amsfonts}

\usepackage{algorithm}
\usepackage{algpseudocode}
\makeatletter 
\newcommand{\linebreakand}{%
  \end{@IEEEauthorhalign}
  \hfill\mbox{}\par
  \mbox{}\hfill\begin{@IEEEauthorhalign}
}
\makeatother 

\usepackage{graphicx}
\usepackage{textcomp}
\usepackage{xcolor}
\usepackage{hyperref}
\def\BibTeX{{\rm B\kern-.05em{\sc i\kern-.025em b}\kern-.08em
    T\kern-.1667em\lower.7ex\hbox{E}\kern-.125emX}}
\begin{document}

\title{Uncertainty Aware Deep Learning Model for Secure and Trustworthy Channel Estimation in 5G Networks}


\author{
\IEEEauthorblockN{Ferhat Ozgur Catak}
\IEEEauthorblockA{\textit{Dept. of Electrical Engineering and Computer Science} \\
\textit{University of Stavanger}\\
Stavanger, Norway \\
f.ozgur.catak@uis.no}
\and
\IEEEauthorblockN{Umit Cali}
\IEEEauthorblockA{Department of Electric Power Engineering \\ 
\textit{Norwegian University of Science and Technology}
\\
Trondheim, Norway \\
umit.cali@ntnu.no}
\and
\linebreakand

\IEEEauthorblockN{Murat Kuzlu}
\IEEEauthorblockA{\textit{Electrical Engineering Technology} \\
\textit{Old Dominion University}\\
Norfolk, VA, USA \\
mkuzlu@odu.edu}
\and
\IEEEauthorblockN{Salih Sarp}
\IEEEauthorblockA{\textit{Electrical and Computer Engineering} \\
\textit{Virginia Commonwealth University}\\
Richmond, VA, USA \\
sarps@vcu.edu}
}


\maketitle

\begin{abstract}
With the rise of intelligent applications, such as self-driving cars and augmented reality, the security and reliability of wireless communication systems have become increasingly crucial. One of the most critical components of ensuring a high-quality experience is channel estimation, which is fundamental for efficient transmission and interference management in wireless networks. However, using deep neural networks (DNNs) in channel estimation raises security and trust concerns due to their complexity and the need for more transparency in decision-making. This paper proposes a Monte Carlo Dropout (MCDO)-based approach for secure and trustworthy channel estimation in 5G networks. Our approach combines the advantages of traditional and deep learning techniques by incorporating conventional pilot-based channel estimation as a prior in the deep learning model. Additionally, we use MCDO to obtain uncertainty-aware predictions, enhancing the model's security and trustworthiness. Our experiments demonstrate that our proposed approach outperforms traditional and deep learning-based approaches regarding security, trustworthiness, and performance in 5G scenarios.
\end{abstract}


\section{Introduction}

Over the past few decades, communication systems have undergone a significant transformation driven by advances in information and processing technologies. Next-generation communication systems, such as 5G and beyond, have garnered significant attention due to their potential to meet the growing demands of industry and consumers. These requirements include higher data transmission speeds, ultra-low latency, increased reliability, a better quality of service, massive network coverage, and increased availability. However, with the proliferation of new applications, such as Augmented Reality (AR)/Virtual Reality (VR), the Internet of Things (IoT), and autonomous vehicles, come significant cyber security risks \cite{iyer2022survey}. These risks stem from the exponential growth in traffic volume and the increased complexity of the communication networks, which seriously threaten the confidentiality, integrity, and availability of data. For instance, an attacker may exploit a vulnerability in the communication network to intercept sensitive information or launch a Distributed Denial of Service (DDoS) attack, which can result in a significant disruption of service. In addition, the ITU-R study on "IMT traffic predictions for the years 2020 to 2030" predicts that mobile traffic (excluding M2M traffic) will expand at an annual rate of roughly 54\% in 2020-2030, reaching 543EB in 2025 and 4394EB in 2030 \cite{union2015imt}. Therefore, it is crucial to develop secure and trustworthy communication systems that can meet the growing demands of industry and consumers while mitigating cyber security risks.

In recent years, NextG technologies have been equipped with improved information, and computing methodologies \cite{10008604}. Various studies have proposed different algorithms for channel estimation. This paper presents a novel approach for channel estimation that employs deep learning models. Channel estimation is a challenging task in wireless communication due to algorithms' high computational complexity level, many mathematical operations, and the low accuracy of channel estimation. To overcome these challenges, machine learning approaches have been successfully used in 5G and beyond communication systems, resulting in improved performance of channel estimation algorithms \cite{simeone2018very}. Deep learning-based algorithms, in particular, offer the potential for reduced computing costs and excellent channel estimation accuracy. However, the trustworthiness of using machine learning methods in NextG technologies is a concern. The lack of transparency in deep learning algorithms makes interpreting and understanding their decision-making processes challenging, leading to uncertainty and trust issues \cite{li2103interpretable}. To address this, research has been conducted on interpretable deep learning models and proposed a taxonomy for categorizing interpretation algorithms. Trustworthy autonomy for NextG technologies, along with explainable AI (XAI) and its test protocols for integration with radio resource management and associated key performance indicators (KPIs) for trust, has also been introduced \cite{9210233}. Building trust and transparency into AI optimization algorithms from the design through the testing phase is crucial for ensuring their trustworthiness in NextG technologies.

\section{Related Work}\label{sec:related}
Quantifying uncertainty in deep learning (DL) models is a topic of ongoing research. One of the primary methods for assessing uncertainty in DL models is using the softmax variance and expected entropy across multiple models. Gal et al. \cite{gal2016uncertainty} showed that a neural network model with dropout activated during prediction time is equivalent to a specific variation of a Bayesian neural network model. This approach approximates the model's uncertainty by averaging probabilistic feed-forward MC dropout sampling during prediction time. This method is efficient for large models and helps understand how networks perform under different test conditions \cite{verdoja2020notes}.

An alternative method for quantifying uncertainty in DL models is through the use of Deep Ensembles, which is a non-Bayesian approach \cite{10.5555/3295222.3295387}. This method trains multiple models in parallel, each with random noise and adversarial instances in its training dataset subset. This results in a set of independent classifiers, each with a unique weight. The ensemble can produce different predictions for the same input instance using its models. However, since Deep Ensembles require multiple models, it is unsuitable for single-model uncertainty quantification. In this study, we have chosen the prediction time-activated dropout-based neural network approach \cite{10.1145/3417330} for single-model uncertainty quantification.

The goal of methods in adversarial ML is to create modified input instances that can evade or fool a DL model. Recently, Ma et al., in their work \cite{10.1145/3417330}, have shown the relationship between uncertainty and model prediction performance. Their approach focused on increasing the robustness of DL models by applying various adversarial ML attacks, such as the Fast Gradient Sign Method, Basic Iterative Method, DeepFool, Jacobian-Based Saliency Map Attack, and Carlini-Wagner, to create instances that are highly uncertain and misclassified. These newly generated instances are then used as a new training dataset in a re-training phase to enhance the DL model's robustness. This iterative training with adversarial instances is similar to traditional adversarial training techniques \cite{2014arXiv1412.6572G}. They evaluated their approach on the MNIST, Fashion MNIST, and CIFAR-10 image datasets. While the method is similar to adversarial training, it improves the prediction performance of DL models by using highly uncertain inputs instead of adversarial ones.

The authors of \cite{ghoshal2020estimating} investigated the use of Bayesian Convolutional Neural Networks (CNN) for quantifying uncertainty in Deep Learning models to improve the diagnostic capabilities of human-machine collaboration using a dataset of COVID-19 chest X-ray images. Another research paper, \cite{schubert2020metadetect}, proposed a method for estimating predictive uncertainty in DL models by considering uncertainty values and quality estimates. However, this approach is limited to object detection tasks and cannot be applied to data from cyber-physical systems (CPS).

The study \cite{zhang2020artificial} provides a comprehensive review of AI-based techniques at the algorithm, implementation, and optimization levels. It highlights the advantages and disadvantages of these solutions, outlines upcoming techniques, and discusses unresolved research problems. Additionally, it introduces machine learning (ML) techniques that handle model uncertainties, such as AI-based full-duplex SI cancellation and RF/PA linearization.  The emergence of network slicing in 5G networks has created a need for new resource allocation algorithms that can satisfy service level agreements (SLAs) and optimize resource utilization while minimizing costs for tenants. The authors in \cite{abozariba2020uncertainty} propose a family of online algorithms that can improve prediction-based decisions and adapt to prediction error variance, thus addressing the issue of uncertainty in prediction models. Results indicate that the proposed probabilistic approach is capable of addressing the uncertainty inherent in prediction-based decisions, thereby improving the overall performance of network resource allocation algorithms for 5G network slicing.


In our recent study \cite{aitest-ieee}, we introduced a new method for measuring uncertainty in object detection models called PURE. To evaluate the effectiveness of PURE, we conducted experiments on a variety of datasets, including KITTI, Stanford cars, Berkeley DeepDrive, and NEXET, using popular object detection models such as YoLo, SSD300, and SSD512. The results showed that PURE effectively quantified the prediction uncertainty of these models.

A variety of methods have been proposed in the literature for addressing uncertainty in requirements, design models, and other areas of software engineering, including RELAX \cite{RELAX}, Fuzzy RE \cite{baresi2010fuzzy}, U-RUCM \cite{U-RUCM}, and UncerTum \cite{UncerTum}. Additionally, some approaches focus on refining uncertainty measurements using historical data, such as the work by Zhang et al. \cite{zhang2017uncertainty} or using Bayesian inference, as in the study by Valdes et al. \cite{valdes2019bayesian}. There also exist methods for testing software systems under uncertainty, such as the approaches by Ma et al. \cite{ma2021testing} and Camilli et al. \cite{camilli2018online}.

Various approaches have been proposed in the literature to ensure the quality and reliability of software systems that employ Deep Learning (DL) models. Zhang~\cite{9270327} proposed a method to enhance the prediction performance of DL models by using adversarial machine learning (ML) attacks to generate highly uncertain test inputs. This approach uses genetic algorithms to create adversarial instances that can effectively penetrate DL defense techniques and reveal hidden defects in the models. The models are then re-trained with these adversarial inputs. On the other hand, Tuna et al.~\cite{2020arXiv201206390F} demonstrated that adversarial inputs generated by adversarial ML attacks could deceive DL models with low prediction uncertainty values. 

In autonomous driving, Michelmore et al. \cite{DBLP:journals/corr/abs-1811-06817} applied variation ratios, entropy, and mutual information-based uncertainty quantification methods to discover the relation between uncertainty and wrong prediction of steering angles of autonomous cars. They found that wrong angle decisions of an autonomous car have significantly higher uncertainty values than correct decisions.

\section{Preliminaries}

\subsection{Uncertainty in Deep Learning}

In their work, Gal et al. in their work \cite{gal2016uncertainty} presented an MC Dropout technique, which incorporates dropout layers in existing object recognition models without altering the number of neurons or layers. This is an ensemble technique in which the system randomly drops out different neurons in each layer during the prediction phase according to a specified dropout ratio. The final prediction is obtained by averaging the predictions from multiple dropout iterations, as represented by the following equation:

\begin{equation}
p(\hat{y}=c|\mathbf{x},\mathcal{D}) \approx \hat{\mu}{pred} = \frac{1}{T} \sum{y \in T} p(\hat{y}|\mathbf{\theta},\mathcal{D})
\end{equation}

Where $\theta$ represents the model's weights, $\mathcal{D}$ is the input dataset, $T$ is the number of predictions of the MC dropouts, and $\mathbf{x}$ is the input sample. The overall prediction uncertainty is approximated by finding the entropy and the probabilistic feed-forward MC dropout sampling variance during prediction time. Using this approach, the class label of the input sample can be estimated by taking the mean value of multiple MC dropouts predictions. This allows for the prediction and quantification of the overall prediction uncertainty.

\subsection{Uncertainty Sources}
DNN models are known to be affected by two types of uncertainties: epistemic \cite{6298890}, and aleatoric \cite{8569814}. Epistemic uncertainty refers to the uncertainty that arises from the model's lack of knowledge. This can be mitigated by increasing the training data and fine-tuning the model's architecture, such as the number of layers and activation functions, to better suit the problem at hand \cite{gal2016uncertainty}. On the other hand, aleatoric uncertainty arises from inherent noise in the data, such as measurement errors. This type of uncertainty cannot be reduced, even with more data. This is an important consideration when working with neural networks, as it can affect the accuracy of the predictions (cite: Kendall and Gal \cite{8569814}).

\subsection{Uncertainty Quantification Methods}

A typical DNN model generates a single output for a regression problem or a softmax probability for each label in a classification problem. To quantify the uncertainty of a single prediction, the model should produce multiple outputs for a single input instance. An uncertainty quantification metric is a scalar value that measures the reliability of the prediction for an input instance. For regression models, the only uncertainty quantification metric is the variance of the predictions:

\begin{equation}
S^2 = \frac{\sum_{i=1}^T(y_i - \overline{y})}{T-1}
\end{equation}

where $y_i$ is the $i$th prediction of the input instance $\mathbf{x}$, and $\overline{y}$ is the average of all the predictions, $y = [y_1, \cdots, t_T]$. 

In addition to variance, several other uncertainty quantification metrics can be used for DNN models. One such metric is the Confidence Interval, which can be computed using the following equation:

\begin{equation}
CI = \overline{y} \pm z_{\alpha/2} \frac{S}{\sqrt{T}}
\end{equation}

where $\overline{y}$ is the mean prediction, $S$ is the standard deviation of the predictions, $T$ is the number of predictions, and $z_{\alpha/2}$ is the critical value of the standard normal distribution for a given confidence level $\alpha$.

Another metric is Predictive Entropy, which is a measure of the Uncertainty of a classifier's prediction, computed as the EntropyEntropy of the predicted class probabilities:

\begin{equation}
H(p) = -\sum_{c=1}^C p(c) \log p(c)
\end{equation}

where $C$ is the number of classes and $p(c)$ is the predicted probability of class $c$.

Additionally, Aleatoric Uncertainty, a measure of the inherent noise in the data, can be estimated by adding a noise term to the model's output. Finally, Epistemic Uncertainty, a measure of Uncertainty due to limited data or model capacity, can be estimated by training an ensemble of models or using Bayesian neural networks. These are just a few examples of uncertainty quantification metrics, and many other methods can be used depending on the specific problem and data set.

\subsection{Channel Estimation}

Without a doubt, one of the most important procedures in wireless communication is channel estimation, which assures the communication channel's performance is high. It is also used to boost overall system performance by enhancing capacity and data rates.
The properties of the link between a transmitter and receiver in a wireless communication system is known as the channel characteristic or channel state information (CSI). When a signal is transmitted through a medium, it can become distorted and contain noise by the time it is received. To properly decode the received signal, it's important to first identify the channel characteristic through a process called channel estimation and remove any unwanted distortion or noise added by the channel. The received signal is attenuated by a factor \(h_0\) and delayed by a specific time \(\tau_0\) which is effected by the propagation speed in communication medium. \(h_0\) also determined by the  $Tx/Rx$ gains, propagation medium and frequency.

Transmitted signal ($x(t)$) is distorted during its propagation through a communication channel, i.e., air, and absorbs noise. That's why received signal (\(y(t)\)) differs from the transmitted signal. Received signal \(y(t)\)  can be represented as:

\begin{equation}
    y(t) = h_0 * x(t - \tau_0)
\end{equation}

However, the received signal is composed of multiple paths, such as reflected and scattered paths, each with its own unique attenuation and delay. This results in a composite received signal.

\begin{equation}
    y(t)=\sum_{l = 0}^{l}h_l * x(t - \tau_l)
\end{equation}
where \(l\) indicates the specific path/tap at a time. 

In addition to composite received signal, the movement of the transmission elements can introduce Doppler effect. This results in the frequency shift which alters the perception of an observer who is moving alongside the signal source.
Channel characteristics, represented by \(h_0\) and \(\tau_0\) are effected greatly by the Doppler effect while calculating multi-path signal loss and fading. The Doppler effected channel characteristics , {\(h_l^t\)} and {\(\tau_l^t\)} incorporates number of paths and the delay. The received signal can be represented as follows:


\begin{equation}
    y(t)=\sum_{l = 0}^{l}h_l^t * x(t - \tau_l^t)
\end{equation}

Channel estimation is a crucial aspect of wireless communications, as it plays a vital role in increasing capacity and overall system performance. 
New technologies are integrated to the communication networks to improve quality of service by providing higher data and network capacity.
Transition from Single Input Single Output (SISO) antennas to Multiple Input Multiple Output (MIMO) antennas is one of these technologies that enables the next-generation (NextG) communication systems. Reliable estimation of channel parameters draws enormous attention as it is the main driver of the NextG systems. 
Channel estimation is executed by merging transmitted signal with pilots which gives the distortion information of the transmitted signal during the transmission. This information is then used to increase receiver's  ability to correctly pick up the original signal.
There are three main categories of channel estimation algorithms, i.e., blind channel estimation, semi-blind channel estimation, and most used training-based estimation method \cite{oyerinde2012review}. 

\section{System Model}
\subsection{Dataset}

The data used in this study benefits the MATLAB 5G Toolbox which provides wide range of NextG communication system samples. This tool support creation and customization of diverse communication modes, subcarrier spacing, number of subcarriers, waveforms, frequency, and channel models.
The example from this toolbox, "Deep Learning Data Synthesis for 5G Channel Estimation", was used to generate the dataset for channel parameters in this study. Estimation of channel parameters are carried out by SISO benefitting CNN models. Channel estimation model is built using the PDSCH and DM-RS symbols. 
The dataset consists of 256 training examples, each with 8568 data points converted into 4-D arrays for use in the CNN model, and is generated using various channel parameters such as delay profiles, delay spreads, doppler shifts and SNR changes. 
The training dataset is created by generating new channel characteristics utilizing diverse tapped delay line (TPL) profiles, delay spreads, SNR values and doppler shifts. These parameters are provided with their respective values in Table \ref{tab:dataset}. Perfect channel values and the DM-RS-based corresponding generated transmitted waveform are included in the dataset.


\begin{table}[!htbp]
\centering
\caption{Each of the channel characteristic parameters and values}
\label{tab:datasetParameter}
\begin{tabular}{|l|c|}
\hline
\textbf{Channel Parameter}& {\textbf{Value }}  \\
 \hline 
Delay Profile & TDL-A, TDL-B, TDL-C, TDL-D, TDL-E \\ \hline
Delay Spread & 1-300 ns \\ \hline
Maximum Doppler Shift & 5-400 Hz\\  \hline
NFFT& 1024  \\ \hline
Sample Rate& 30720000  \\  \hline
Symbols Per Slot& 14 \\  \hline
Windowing& 36 \\  \hline
Slots Per Subframe& 2 \\ \hline
Slots Per Frame& 20 \\ \hline
Polarization & Co-Polar \\ \hline
Transmission Direction & Downlink \\ \hline
Num. Transmit Antennas & 1 \\ \hline
Num. Receive Antennas & 1 \\ \hline
Fading Distribution & Rayleigh\\ \hline
Modulation & 16QAM\\ \hline
\end{tabular}
\label{tab:dataset}
\end{table}

In order to prevent overfitting, the training dataset is divided into two sets: a validation set and a training set. Model training is done with the training set (80\% of the dataset) whereas validation set (20\% of the dataset) is utilized to evaluate how well neural network learnt the data distribution. 

\subsection{Approach}

In this paper, we present a novel approach for channel estimation using a combination of traditional and deep learning techniques to improve the trustworthiness of next-generation communication systems. Specifically, our approach incorporates a Monte-Carlo Dropouts (MCDO) technique for uncertainty-aware prediction, which enhances the reliability of deep learning-based algorithms \cite{gal2016uncertainty}. Moreover, we integrate the conventional pilot-based channel estimation technique as a prior in our deep learning model to leverage the advantages of both traditional and deep learning methods. This combination of techniques not only improves the accuracy of channel estimation but also enhances the transparency and interpretability of the resulting model, which is essential for ensuring the trustworthiness of our approach in practical applications.


The proposed uncertainty-aware trustworthy deep learning approach for channel estimation combines the advantages of both traditional and deep learning techniques. Our method is based on MCDO technique for uncertainty-aware prediction. Additionally, our approach leverages the traditional channel estimation framework by incorporating the conventional pilot-based channel estimation as a prior in the deep learning model. To improve the model's trustworthiness, we adopted an adversarial retraining approach for the MCDO technique. Adversarial retraining uses an adversarial network to generate negative examples for the training set and then retrains the models with the generated adversarial examples. In our approach, we selected highly uncertain inputs to retrain the model iteratively. To compute the uncertainty of the model, we used the uncertainty wizard library. Algorithm \ref{alg:uncertainty_aware_trustworthy_deep_learning} presents the proposed approach for channel estimation DL model retraining using highly uncertain inputs.

\begin{algorithm}[!htbp]
\caption{Uncertainty Aware Trustworthy Deep Learning Approach for Channel Estimation.}
\label{alg:uncertainty_aware_trustworthy_deep_learning} \small
\begin{algorithmic}[1]
\State \textbf{Inputs:} DL model, training set $\mathcal{D}$, validation set $\mathcal{V}$, max iterations $T$, tolerance $\epsilon$
\State \textbf{Output:} Trained trustworthy DL model $M$
\State Initialize DL model $M$
\For{$t \gets 1$ to $T$}
    \State Compute uncertainty of the model $M$ on the validation set $\mathcal{V}$
    \State Select highly uncertain inputs from the validation set $\mathcal{V}$
    \State Generate adversarial examples from the selected inputs
    \State Retrain the DL model $M$ with $\mathcal{D} \cup \mathcal{V}$
    \State Compute the loss of the DL model $M$ on the validation set $\mathcal{V}$
    \If{loss $< \epsilon$}
        \State \textbf{break}
    \EndIf
\EndFor
\State \textbf{return} Trained trustworthy DL model $M$
\end{algorithmic}
\end{algorithm}

Algorithm \ref{alg:uncertainty_aware_trustworthy_deep_learning} is performed by using the uncertainty wizard library \cite{Weiss2021FailSafe}. The library provides several uncertainty metrics, such as entropy, mutual information, and prediction variance. In this paper, we used entropy as the uncertainty metric.

\begin{figure}[!htbp]
    \centering
    \includegraphics[width=1.0\linewidth]{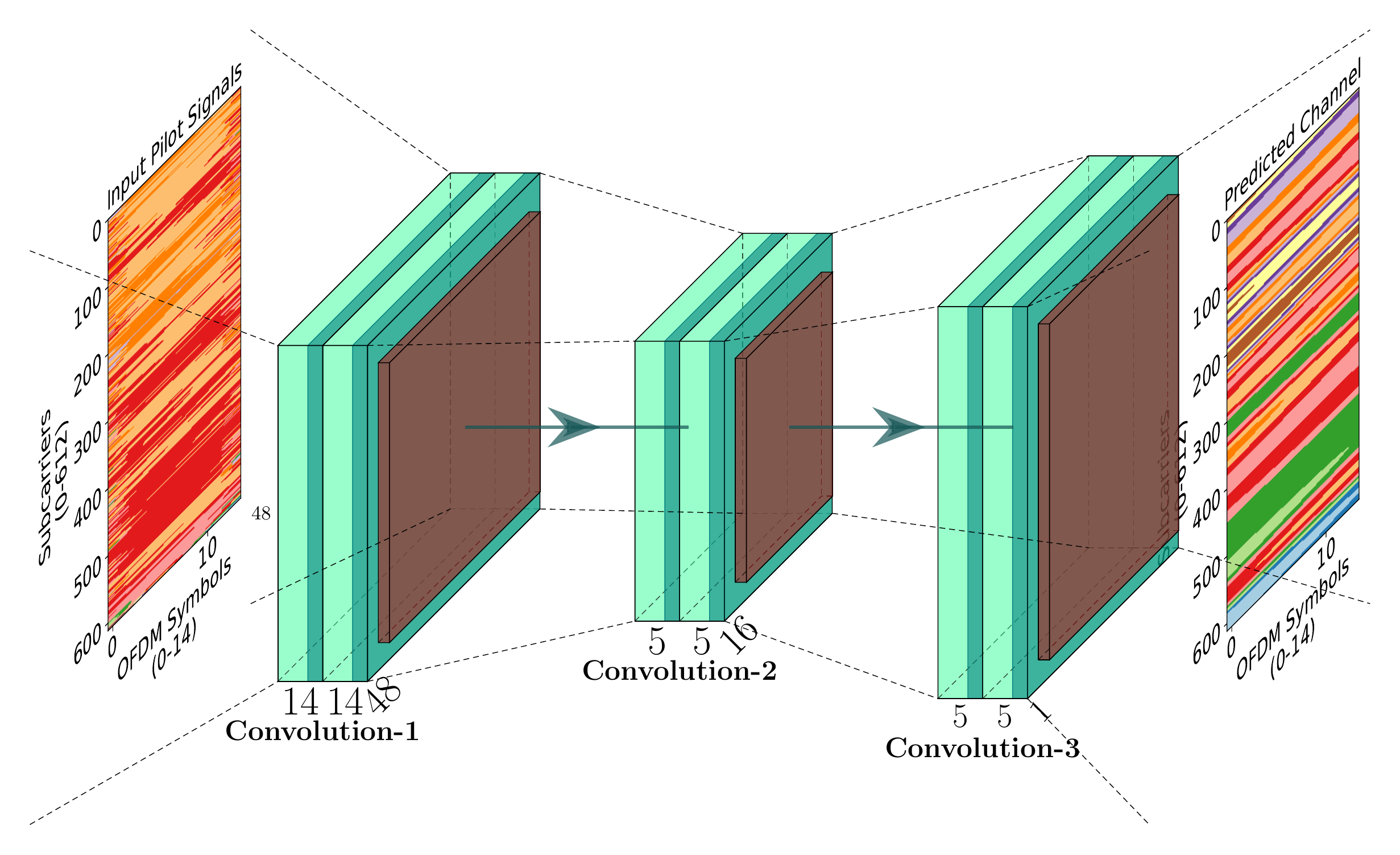}
    \caption{DL Model}
    \label{fig:model_plot}
\end{figure}

\section{Experiments and Results}
\label{sec:experiments}

\subsection{Experiment Setup}

We conducted experiments to evaluate the performance and trustworthiness of our proposed approach for channel estimation in 5G scenarios. Our proposed approach was compared to the traditional and deep learning-based approaches. The conventional approach is based on pilot-based channel estimation, while the deep learning-based process is a CNN model. All the models were implemented using the Python programming language, and the TensorFlow framework.

The dataset for the experiments was generated using the MATLAB 5G Toolbox. The parameters of the channel estimation scenario are given in Table \ref{tab:datasetParameter}. The models were trained with the generated dataset using the Adam optimizer. The training was stopped when the validation loss stopped decreasing and improving the model. The learning rate was set to 0.001, and the batch size was set to 64.

\subsection{Results and Discussion}

\begin{figure}[!htbp]
    \centering
    \includegraphics[width=0.8\linewidth]{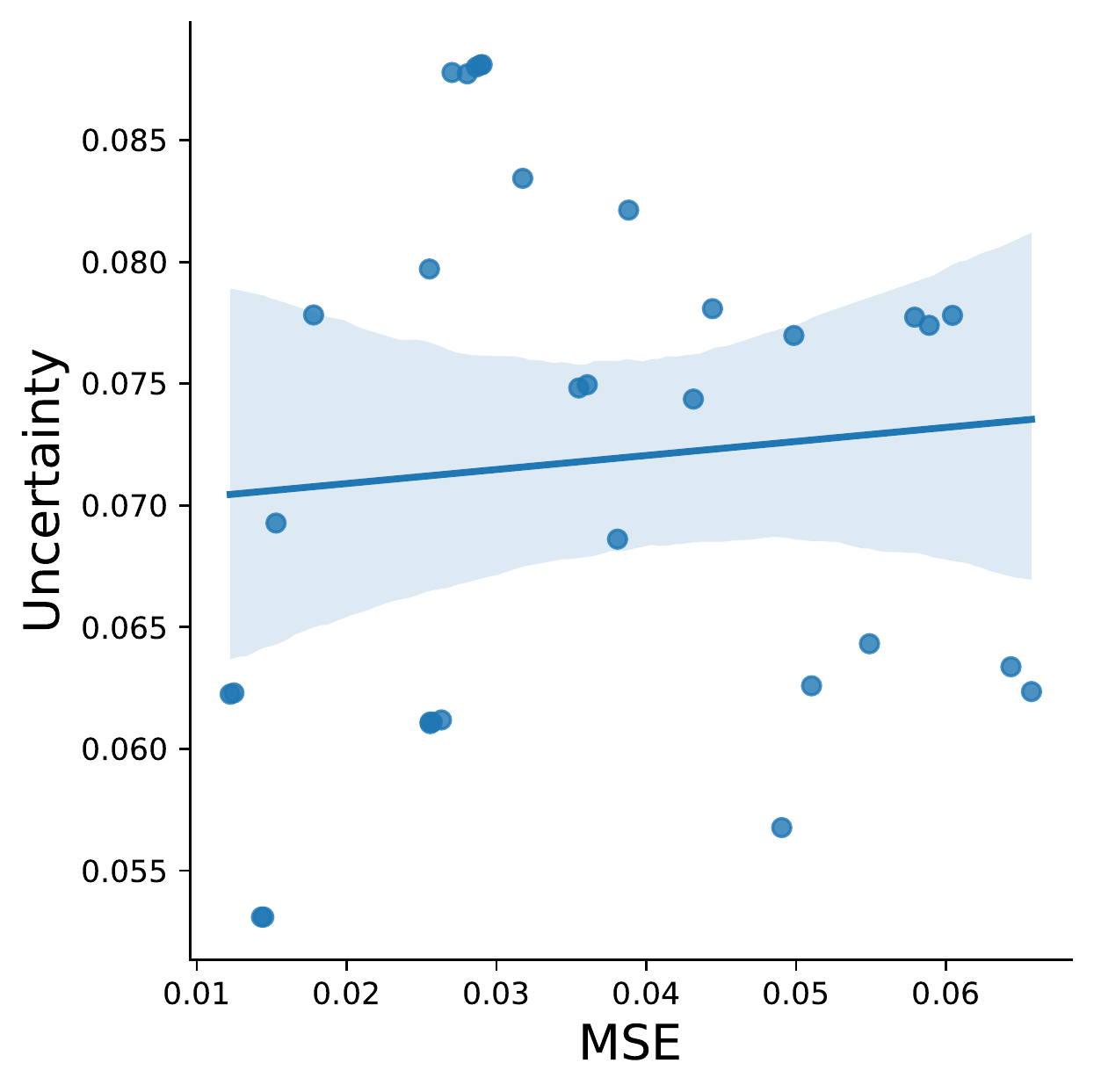}
    \caption{Initial MSE and uncertainty correlations of the traditional and deep learning-based models.}
    \label{fig:MSE}
\end{figure}

Figure \ref{fig:MSE} shows the initial mean squared error (MSE) and Uncertainty correlations of the traditional and deep learning-based models. It can be seen that the deep learning-based model has a lower MSE compared to the conventional model, which indicates better channel estimation accuracy. This is because the deep learning-based model can extract more information from the received signal than the traditional model.

\begin{figure}[!htbp]
    \centering
    \includegraphics[width=0.8\linewidth]{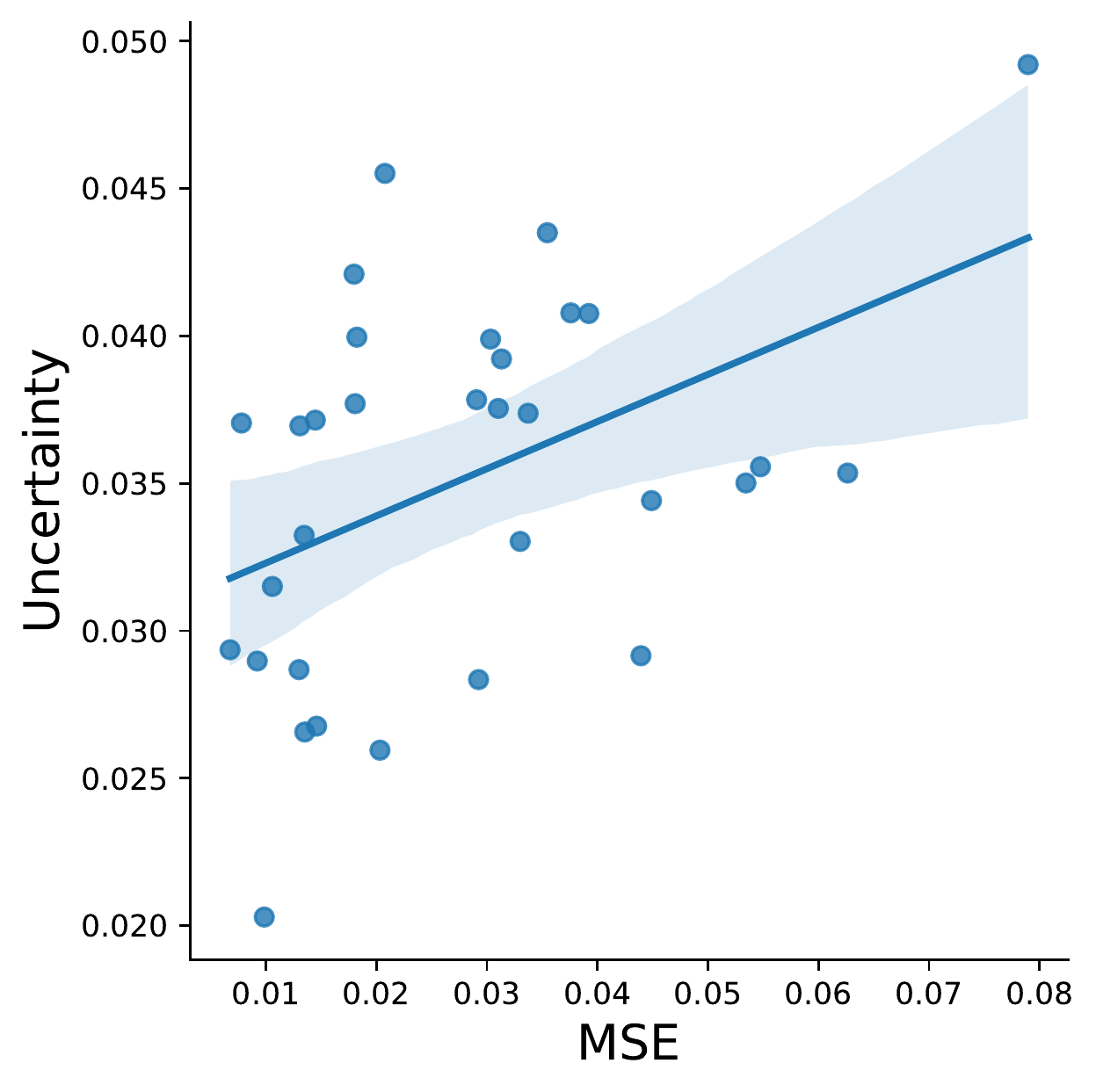}
    \caption{Retrained MSE and uncertainty correlations of the DL model.}
    \label{fig:retrained_MSE}
\end{figure}

Figure \ref{fig:retrained_MSE} shows the MSE and uncertainty correlations of the retrained DL model. It can be seen that the retrained DL model has lower MSE compared to the initial DL model, which indicates better channel estimation accuracy. This is because the retrained DL model can exploit the uncertainties in the validation set to improve the model's performance. Furthermore, the uncertainty correlations are enhanced after the retraining, which indicates that the DL model is more trustworthy.

\begin{figure}[!htbp]
    \centering
    \includegraphics[width=0.9\linewidth]{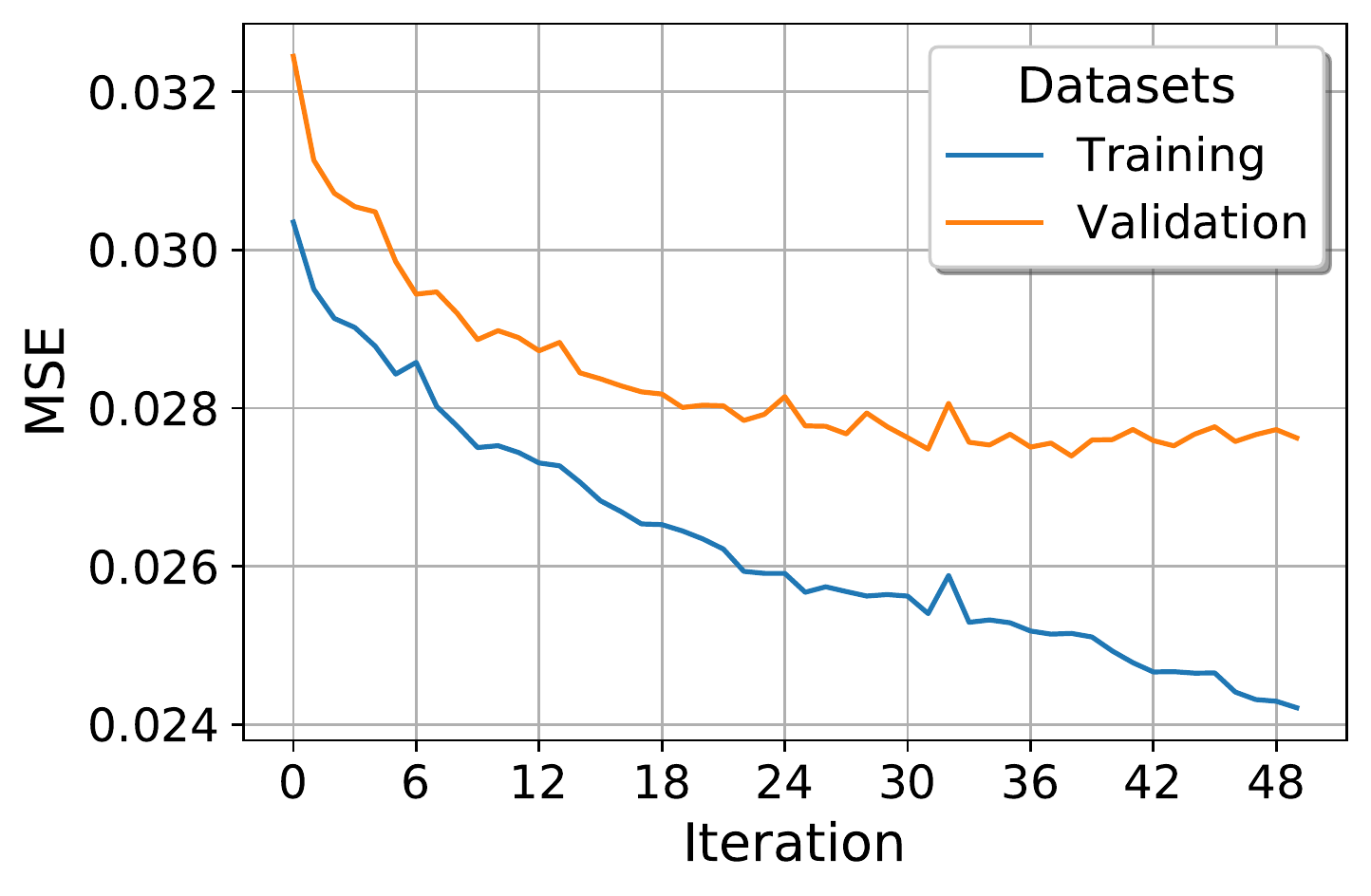}
    \caption{MSE improvements in each iteration of the retrained DL model.}
    \label{fig:comparison}
\end{figure}

Figure \ref{fig:comparison} shows MSE improvements in each iteration for training and validation sets of the retrained DL model. It can be seen that the MSE is improved with each iteration in both training and validation sets. This indicates that our proposed approach is able to improve the channel estimation accuracy of the DL model. 

\section{Conclusion and Future Work}

NextG technologies, such as 5G and beyond, have received significant attention from the industry and academia to meet the growing demand for high-speed data transmission, low latency, and increased reliability. Advanced information and computing methods, such as channel estimation and machine learning, have been applied to NextG technologies to achieve these requirements. However, the trustworthiness of using machine learning methods in NextG technologies remains a crucial concern.

In this study, we proposed an uncertainty-aware trustworthy deep learning approach for channel estimation that combines the benefits of traditional and deep learning techniques. Our approach leverages the traditional pilot-based channel estimation framework as a prior in the deep learning model and uses MCDO technique for uncertainty-aware prediction. Additionally, we adopted an adversarial retraining approach to improve the model's trustworthiness by selecting highly uncertain inputs for iterative retraining. The experimental results showed that our approach outperforms conventional and deep learning-based approaches in performance and trustworthiness.

In future work, we plan to investigate the generalization capability of our proposed approach to different channel environments and evaluate its performance in various scenarios. We also aim to explore alternative techniques for uncertainty estimation and investigate their effectiveness in improving the trustworthiness of deep learning-based channel estimation. Moreover, we will consider further integrating our approach with other communication systems components, such as resource allocation and power control, to enhance the overall system performance. Ultimately, the proposed approach can contribute to developing more trustworthy and reliable NextG communication systems.

\bibliographystyle{IEEEtran}
\bibliography{refs.bib}

\begin{thebibliography}{10}

\bibitem{iyer2022survey}
Sridhar Iyer, Anita Patil, Shilpa Bhairanatti, Soumya Halagatti, and
  Rahul~Jashvantbhai Pandya.
\newblock A survey on technological trends to enhance spectrum-efficiency in 6g
  communications.
\newblock {\em Transactions of the Indian National Academy of Engineering},
  7(4):1093--1120, 2022.

\bibitem{union2015imt}
IIMT Union.
\newblock Imt traffic estimates for the years 2020 to 2030.
\newblock {\em Report ITU}, 2370, 2015.

\bibitem{10008604}
Ferhat~Ozgur Catak, Murat Kuzlu, Salih Sarp, Evren Catak, and Umit Cali.
\newblock Mitigating attacks on artificial intelligence-based spectrum sensing
  for cellular network signals.
\newblock In {\em 2022 IEEE Globecom Workshops (GC Wkshps)}, pages 1371--1376,
  2022.

\bibitem{simeone2018very}
Osvaldo Simeone.
\newblock A very brief introduction to machine learning with applications to
  communication systems.
\newblock {\em IEEE Transactions on Cognitive Communications and Networking},
  4(4):648--664, 2018.

\bibitem{li2103interpretable}
X~Li, H~Xiong, X~Li, X~Wu, X~Zhang, J~Liu, J~Bian, and D~Dou.
\newblock Interpretable deep learning: Interpretation, interpretability,
  trustworthiness, and beyond. arxiv 2021.
\newblock {\em arXiv preprint arXiv:2103.10689}.

\bibitem{9210233}
Chen Li, Weisi Guo, Schyler~Chengyao Sun, Saba Al-Rubaye, and Antonios
  Tsourdos.
\newblock Trustworthy deep learning in 6g-enabled mass autonomy: From concept
  to quality-of-trust key performance indicators.
\newblock {\em IEEE Vehicular Technology Magazine}, 15(4):112--121, 2020.

\bibitem{gal2016uncertainty}
Yarin Gal.
\newblock Uncertainty in deep learning.
\newblock {\em University of Cambridge}, 1(3), 2016.

\bibitem{verdoja2020notes}
Francesco Verdoja and Ville Kyrki.
\newblock Notes on the behavior of mc dropout, 2020.

\bibitem{10.5555/3295222.3295387}
Balaji Lakshminarayanan, Alexander Pritzel, and Charles Blundell.
\newblock Simple and scalable predictive uncertainty estimation using deep
  ensembles.
\newblock In {\em Proceedings of the 31st International Conference on Neural
  Information Processing Systems}, NIPS'17, page 6405–6416, Red Hook, NY,
  USA, 2017. Curran Associates Inc.

\bibitem{10.1145/3417330}
Wei Ma, Mike Papadakis, Anestis Tsakmalis, Maxime Cordy, and Yves~Le Traon.
\newblock Test selection for deep learning systems.
\newblock {\em ACM Trans. Softw. Eng. Methodol.}, 30(2), January 2021.

\bibitem{2014arXiv1412.6572G}
Ian~J. {Goodfellow}, Jonathon {Shlens}, and Christian {Szegedy}.
\newblock {Explaining and Harnessing Adversarial Examples}.
\newblock {\em arXiv e-prints}, 1(1):arXiv:1412.6572, December 2014.

\bibitem{ghoshal2020estimating}
Biraja Ghoshal and Allan Tucker.
\newblock Estimating uncertainty and interpretability in deep learning for
  coronavirus (covid-19) detection, 2020.

\bibitem{schubert2020metadetect}
Marius Schubert, Karsten Kahl, and Matthias Rottmann.
\newblock Metadetect: Uncertainty quantification and prediction quality
  estimates for object detection, 2020.

\bibitem{zhang2020artificial}
Chuan Zhang, Yeong-Luh Ueng, Christoph Studer, and Andreas Burg.
\newblock Artificial intelligence for 5g and beyond 5g: Implementations,
  algorithms, and optimizations.
\newblock {\em IEEE Journal on Emerging and Selected Topics in Circuits and
  Systems}, 10(2):149--163, 2020.

\bibitem{abozariba2020uncertainty}
Raouf Abozariba, Muhammad~Kamran Naeem, Md~Asaduzzaman, and Mohammad Patwary.
\newblock Uncertainty-aware ran slicing via machine learning predictions in
  next-generation networks.
\newblock In {\em 2020 IEEE 92nd Vehicular Technology Conference
  (VTC2020-Fall)}, pages 1--6. IEEE, 2020.

\bibitem{aitest-ieee}
Ferhat~Ozgur Catak, Tao Yue, and Shaukat Ali.
\newblock Prediction surface uncertainty quantification in object detection
  models for autonomous driving.
\newblock In {\em 2021 IEEE International Conference on Artificial Intelligence
  Testing (AITest)}, pages 93--100, 2021.

\bibitem{RELAX}
Betty~HC Cheng, Pete Sawyer, Nelly Bencomo, and Jon Whittle.
\newblock A goal-based modeling approach to develop requirements of an adaptive
  system with environmental uncertainty.
\newblock In {\em International Conference on Model Driven Engineering
  Languages and Systems}, pages 468--483. Springer, 2009.

\bibitem{baresi2010fuzzy}
Luciano Baresi, Liliana Pasquale, and Paola Spoletini.
\newblock Fuzzy goals for requirements-driven adaptation.
\newblock In {\em 2010 18th IEEE International Requirements Engineering
  Conference}, pages 125--134. IEEE, 2010.

\bibitem{U-RUCM}
Man Zhang, Tao Yue, Shaukat Ali, Bran Selic, Oscar Okariz, Roland Norgre, and
  Karmele Intxausti.
\newblock Specifying uncertainty in use case models.
\newblock {\em Journal of Systems and Software}, 144:573--603, 2018.

\bibitem{UncerTum}
Man Zhang, Shaukat Ali, Tao Yue, Roland Norgren, and Oscar Okariz.
\newblock Uncertainty-wise cyber-physical system test modeling.
\newblock {\em Software \& Systems Modeling}, 18(2):1379--1418, 2019.

\bibitem{zhang2017uncertainty}
Man Zhang, Shaukat Ali, Tao Yue, and Roland Norgre.
\newblock Uncertainty-wise evolution of test ready models.
\newblock {\em Information and Software Technology}, 87:140--159, 2017.

\bibitem{valdes2019bayesian}
Rosa~Arnaldo Vald{\'e}s, Victor Fernando~Gomez Comendador, Javier Alberto~Perez
  Castan, Alvaro~Rodriguez Sanz, Luis~Perez Sanz, Francisco Javier~Saez Nieto,
  and Eduardo~Sanchez Aira.
\newblock Bayesian inference in safety compliance assessment under conditions
  of uncertainty for ans providers.
\newblock {\em Safety science}, 116:183--195, 2019.

\bibitem{ma2021testing}
Tao Ma, Shaukat Ali, and Tao Yue.
\newblock Testing self-healing cyber-physical systems under uncertainty with
  reinforcement learning: an empirical study.
\newblock {\em Empirical Software Engineering}, 26(3):1--54, 2021.

\bibitem{camilli2018online}
Matteo Camilli, Carlo Bellettini, Angelo Gargantini, and Patrizia Scandurra.
\newblock Online model-based testing under uncertainty.
\newblock In {\em 2018 IEEE 29th International Symposium on Software
  Reliability Engineering (ISSRE)}, pages 36--46. IEEE, 2018.

\bibitem{9270327}
Xiyue Zhang.
\newblock Uncertainty-guided testing and robustness enhancement for deep
  learning systems.
\newblock In {\em 2020 IEEE/ACM 42nd International Conference on Software
  Engineering: Companion Proceedings (ICSE-Companion)}, pages 101--103, 2020.

\bibitem{2020arXiv201206390F}
Omer {Faruk Tuna}, Ferhat {Ozgur Catak}, and M.~{Taner Eskil}.
\newblock Closeness and uncertainty aware adversarial examples detection in
  adversarial machine learning.
\newblock {\em Computers and Electrical Engineering}, 101:107986, 2022.

\bibitem{DBLP:journals/corr/abs-1811-06817}
Rhiannon Michelmore, Marta Kwiatkowska, and Yarin Gal.
\newblock Evaluating uncertainty quantification in end-to-end autonomous
  driving control.
\newblock {\em CoRR}, abs/1811.06817, 2018.

\bibitem{6298890}
Y.~{Li}, J.~{Chen}, and L.~{Feng}.
\newblock Dealing with uncertainty: A survey of theories and practices.
\newblock {\em IEEE Transactions on Knowledge and Data Engineering},
  25(11):2463--2482, 2013.

\bibitem{8569814}
D.~{Feng}, L.~{Rosenbaum}, and K.~{Dietmayer}.
\newblock Towards safe autonomous driving: Capture uncertainty in the deep
  neural network for lidar 3d vehicle detection.
\newblock In {\em 2018 21st International Conference on Intelligent
  Transportation Systems (ITSC)}, pages 3266--3273. IEEE, 2018.

\bibitem{oyerinde2012review}
Olutayo~O Oyerinde and Stanley~H Mneney.
\newblock Review of channel estimation for wireless communication systems.
\newblock {\em IETE Technical review}, 29(4):282--298, 2012.

\bibitem{Weiss2021FailSafe}
Michael Weiss and Paolo Tonella.
\newblock Fail-safe execution of deep learning based systems through
  uncertainty monitoring.
\newblock In {\em 2021 14th IEEE Conference on Software Testing, Verification
  and Validation (ICST)}, pages 24--35. IEEE, 2021.

\end{thebibliography}


@article{viswanathan2020communications,
  title={Communications in the 6G era},
  author={Viswanathan, Harish and Mogensen, Preben E},
  journal={IEEE Access},
  volume={8},
  pages={57063--57074},
  year={2020},
  publisher={IEEE}
}
@article{liu2020adversarial,
  title={Adversarial attack on DL-based massive MIMO CSI feedback},
  author={Liu, Qing and Guo, Jiajia and Wen, Chao-Kai and Jin, Shi},
  journal={Journal of Communications and Networks},
  volume={22},
  number={3},
  pages={230--235},
  year={2020},
  publisher={KICS}
}

@article{roh2014millimeter,
  title={Millimeter-wave beamforming as an enabling technology for 5G cellular communications: Theoretical feasibility and prototype results},
  author={Roh, Wonil and Seol, Ji-Yun and Park, Jeongho and Lee, Byunghwan and Lee, Jaekon and Kim, Yungsoo and Cho, Jaeweon and Cheun, Kyungwhoon and Aryanfar, Farshid},
  journal={IEEE communications magazine},
  volume={52},
  number={2},
  pages={106--113},
  year={2014},
  publisher={IEEE}
}
@inproceedings{yizhan20206g,
  title={6G is coming: Discussion on key candidate technologies and application scenarios},
  author={Yizhan, Chen and Zhong, Wang and Da, Huang and Ruosen, Liao},
  booktitle={2020 International Conference on Computer Communication and Network Security (CCNS)},
  pages={59--62},
  year={2020},
  organization={IEEE}
}
@article{xiao2020toward,
  title={Toward self-learning edge intelligence in 6G},
  author={Xiao, Yong and Shi, Guangming and Li, Yingyu and Saad, Walid and Poor, H Vincent},
  journal={IEEE Communications Magazine},
  volume={58},
  number={12},
  pages={34--40},
  year={2020},
  publisher={IEEE}
}

@article{sim2020deep,
  title={Deep learning-based mmWave beam selection for 5G NR/6G with sub-6 GHz channel information: Algorithms and prototype validation},
  author={Sim, Min Soo and Lim, Yeon-Geun and Park, Sang Hyun and Dai, Linglong and Chae, Chan-Byoung},
  journal={IEEE Access},
  volume={8},
  pages={51634--51646},
  year={2020},
  publisher={IEEE}
}


@article{huang2019survey,
  title={A survey on green 6G network: Architecture and technologies},
  author={Huang, Tongyi and Yang, Wu and Wu, Jun and Ma, Jin and Zhang, Xiaofei and Zhang, Daoyin},
  journal={IEEE access},
  volume={7},
  pages={175758--175768},
  year={2019},
  publisher={IEEE}
}

@article{alkhateeb2018deep,
  title={Deep learning coordinated beamforming for highly-mobile millimeter wave systems},
  author={Alkhateeb, Ahmed and Alex, Sam and Varkey, Paul and Li, Ying and Qu, Qi and Tujkovic, Djordje},
  journal={IEEE Access},
  volume={6},
  pages={37328--37348},
  year={2018},
  publisher={IEEE}
}



@article{jungnickel2014role,
  title={The role of small cells, coordinated multipoint, and massive MIMO in 5G},
  author={Jungnickel, Volker and Manolakis, Konstantinos and Zirwas, Wolfgang and Panzner, Berthold and Braun, Volker and Lossow, Moritz and Sternad, Mikael and Apelfr{\"o}jd, Rikke and Svensson, Tommy},
  journal={IEEE communications magazine},
  volume={52},
  number={5},
  pages={44--51},
  year={2014},
  publisher={IEEE}
}

@article{ccatak2016waveform,
  title={Waveform design considerations for 5G wireless networks},
  author={{\c{C}}atak, Evren and Durak-Ata, L{\"u}tfiye},
  journal={Towards 5G Wireless Networks-A Physical Layer Perspective},
  pages={27--48},
  year={2016},
  publisher={IntechOpen}
}

@ARTICLE{2020arXiv200702617A,
       author = {{Andriushchenko}, Maksym and {Flammarion}, Nicolas},
        title = "{Understanding and Improving Fast Adversarial Training}",
      journal = {arXiv e-prints},
     keywords = {Computer Science - Machine Learning, Computer Science - Cryptography and Security, Computer Science - Computer Vision and Pattern Recognition, Statistics - Machine Learning},
         year = 2020,
        month = jul,
          eid = {arXiv:2007.02617},
        pages = {arXiv:2007.02617},
archivePrefix = {arXiv},
       eprint = {2007.02617},
 primaryClass = {cs.LG},
       adsurl = {https://ui.adsabs.harvard.edu/abs/2020arXiv200702617A},
      adsnote = {Provided by the SAO/NASA Astrophysics Data System}
}

@INPROCEEDINGS{8403769,
  author={Lichtman, Marc and Rao, Raghunandan and Marojevic, Vuk and Reed, Jeffrey and Jover, Roger Piqueras},
  booktitle={2018 IEEE International Conference on Communications Workshops (ICC Workshops)}, 
  title={5{G} {NR} Jamming, Spoofing, and Sniffing: Threat Assessment and Mitigation}, 
  year={2018},
  volume={},
  number={},
  pages={1-6},
  doi={10.1109/ICCW.2018.8403769}}
  
@INPROCEEDINGS{9527756,  author={Catak, Evren and Catak, Ferhat Ozgur and Moldsvor, Arild},  booktitle={2021 IEEE International Black Sea Conference on Communications and Networking (BlackSeaCom)},   title={Adversarial Machine Learning Security Problems for 6{G}: mmWave Beam Prediction Use-Case},   year={2021},  volume={},  number={},  pages={1-6},  doi={10.1109/BlackSeaCom52164.2021.9527756}}


@article{catak2021security,
title = {Security concerns on machine learning solutions for 6{G}  networks in mm{W}ave beam prediction},
journal = {Physical Communication},
pages = {101626},
year = {2022},
issn = {1874-4907},
doi = {https://doi.org/10.1016/j.phycom.2022.101626},
url = {https://www.sciencedirect.com/science/article/pii/S1874490722000155},
author = {Ferhat Ozgur Catak and Murat Kuzlu and Evren Catak and Umit Cali and Devrim Unal},
keywords = {Machine learning, AI, Millimeter-wave (mmWave), Beamforming, Adversarial machine learning, 6G, Deep learning},
}

@ARTICLE{9206115,
  author={Du, Jun and Jiang, Chunxiao and Wang, Jian and Ren, Yong and Debbah, Merouane},
  journal={IEEE Vehicular Technology Magazine}, 
  title={Machine Learning for 6{G} Wireless Networks: Carrying Forward Enhanced Bandwidth, Massive Access, and Ultrareliable/Low-Latency Service}, 
  year={2020},
  volume={15},
  number={4},
  pages={122-134},
  doi={10.1109/MVT.2020.3019650}}
  
  @ARTICLE{9023459,
  author={Gui, Guan and Liu, Miao and Tang, Fengxiao and Kato, Nei and Adachi, Fumiyuki},
  journal={IEEE Wireless Communications}, 
  title={6{G}: Opening New Horizons for Integration of Comfort, Security, and Intelligence}, 
  year={2020},
  volume={27},
  number={5},
  pages={126-132},
  doi={10.1109/MWC.001.1900516}}
  
  @misc{ozpoyraz2022deep,
      title={Deep Learning-Aided 6{G} Wireless Networks: A Comprehensive Survey of Revolutionary PHY Architectures}, 
      author={Burak Ozpoyraz and A. Tugberk Dogukan and Yarkin Gevez and Ufuk Altun and Ertugrul Basar},
      year={2022},
      eprint={2201.03866},
      archivePrefix={arXiv},
      primaryClass={cs.IT}
}

@misc{ali20206g,
      title={6{G} White Paper on Machine Learning in Wireless Communication Networks}, 
      author={Samad Ali and Walid Saad and Nandana Rajatheva and Kapseok Chang and Daniel Steinbach and Benjamin Sliwa and Christian Wietfeld and Kai Mei and Hamid Shiri and Hans-Jürgen Zepernick and Thi My Chinh Chu and Ijaz Ahmad and Jyrki Huusko and Jaakko Suutala and Shubhangi Bhadauria and Vimal Bhatia and Rangeet Mitra and Saidhiraj Amuru and Robert Abbas and Baohua Shao and Michele Capobianco and Guanghui Yu and Maelick Claes and Teemu Karvonen and Mingzhe Chen and Maksym Girnyk and Hassan Malik},
      year={2020},
      eprint={2004.13875},
      archivePrefix={arXiv},
      primaryClass={cs.IT}
}

@article{zhang20196g,
  title={6{G} wireless networks: Vision, requirements, architecture, and key technologies},
  author={Zhang, Zhengquan and Xiao, Yue and Ma, Zheng and Xiao, Ming and Ding, Zhiguo and Lei, Xianfu and Karagiannidis, George K and Fan, Pingzhi},
  journal={IEEE Vehicular Technology Magazine},
  volume={14},
  number={3},
  pages={28--41},
  year={2019},
  publisher={IEEE}
}

@article{giordani2020toward,
  title={Toward 6{G} networks: Use cases and technologies},
  author={Giordani, Marco and Polese, Michele and Mezzavilla, Marco and Rangan, Sundeep and Zorzi, Michele},
  journal={IEEE Communications Magazine},
  volume={58},
  number={3},
  pages={55--61},
  year={2020},
  publisher={IEEE}
}

@article{saad2019vision,
  title={A vision of 6{G} wireless systems: Applications, trends, technologies, and open research problems},
  author={Saad, Walid and Bennis, Mehdi and Chen, Mingzhe},
  journal={IEEE network},
  volume={34},
  number={3},
  pages={134--142},
  year={2019},
  publisher={IEEE}
}

@ARTICLE{9163104,
  author={Khan, Latif U. and Yaqoob, Ibrar and Imran, Muhammad and Han, Zhu and Hong, Choong Seon},
  journal={IEEE Access}, 
  title={6{G} Wireless Systems: A Vision, Architectural Elements, and Future Directions}, 
  year={2020},
  volume={8},
  number={},
  pages={147029-147044},
  doi={10.1109/ACCESS.2020.3015289}}
  
  @article{zheng2021potential,
  title={Potential technologies and applications based on deep learning in the 6{G} networks},
  author={Zheng, Zunxin and Wang, Linmei and Zhu, Fumin and Liu, Ling},
  journal={Computers \& Electrical Engineering},
  volume={95},
  pages={107373},
  year={2021},
  publisher={Elsevier}
}

  @article{kuzlu2021role,
  title={Role of Artificial Intelligence in the Internet of Things ({IoT}) cybersecurity},
  author={Kuzlu, Murat and Fair, Corinne and Guler, Ozgur},
  journal={Discover Internet of Things},
  volume={1},
  number={1},
  pages={1--14},
  year={2021},
  publisher={Springer}
}

@inproceedings{siriwardhana2021ai,
  title={{AI} and 6{G} security: Opportunities and challenges},
  author={Siriwardhana, Yushan and Porambage, Pawani and Liyanage, Madhusanka and Ylianttila, Mika},
  booktitle={Proc. IEEE Joint Eur. Conf. Netw. Commun.(EuCNC) 6{G} Summit},
  pages={1--6},
  year={2021}
}

@ARTICLE{9237460,
  author={Yang, Helin and Alphones, Arokiaswami and Xiong, Zehui and Niyato, Dusit and Zhao, Jun and Wu, Kaishun},
  journal={IEEE Network}, 
  title={Artificial-Intelligence-Enabled Intelligent 6{G} Networks}, 
  year={2020},
  volume={34},
  number={6},
  pages={272-280},
  doi={10.1109/MNET.011.2000195}}
  
  @inproceedings{porambage20216g,
  title={6{G} security challenges and potential solutions},
  author={Porambage, Pawani and G{\"u}r, G{\"u}rkan and Osorio, Diana Pamela Moya and Liyanage, Madhusanka and Ylianttila, Mika},
  booktitle={Proc. IEEE Joint Eur. Conf. Netw. Commun.(EuCNC) 6{G} Summit},
  pages={1--6},
  year={2021}
}

@article{dang2020should,
  title={What should 6{G} be?},
  author={Dang, Shuping and Amin, Osama and Shihada, Basem and Alouini, Mohamed-Slim},
  journal={Nature Electronics},
  volume={3},
  number={1},
  pages={20--29},
  year={2020},
  publisher={Nature Publishing Group}
}

@article{liu20185g,
  title={5{G} features from operation perspective and fundamental performance validation by field trial},
  author={Liu, Guangyi and Huang, Yuhong and Wang, Fei and Liu, Jianjun and Wang, Qixing},
  journal={China Communications},
  volume={15},
  number={11},
  pages={33--50},
  year={2018},
  publisher={IEEE}
}

@article{de2021survey,
  title={Survey on 6{G} frontiers: Trends, applications, requirements, technologies and future research},
  author={De Alwis, Chamitha and Kalla, Anshuman and Pham, Quoc-Viet and Kumar, Pardeep and Dev, Kapal and Hwang, Won-Joo and Liyanage, Madhusanka},
  journal={IEEE Open Journal of the Communications Society},
  volume={2},
  pages={836--886},
  year={2021},
  publisher={IEEE}
}

@article{sheth2020taxonomy,
  title={A taxonomy of {AI} techniques for 6{G} communication networks},
  author={Sheth, Karan and Patel, Keyur and Shah, Het and Tanwar, Sudeep and Gupta, Rajesh and Kumar, Neeraj},
  journal={Computer Communications},
  volume={161},
  pages={279--303},
  year={2020},
  publisher={Elsevier}
}


@misc{hinton2015distilling,
      title={Distilling the Knowledge in a Neural Network}, 
      author={Geoffrey Hinton and Oriol Vinyals and Jeff Dean},
      year={2015},
      eprint={1503.02531},
      archivePrefix={arXiv},
      primaryClass={stat.ML}
}

@misc{papernot2016distillation,
      title={Distillation as a Defense to Adversarial Perturbations against Deep Neural Networks}, 
      author={Nicolas Papernot and Patrick McDaniel and Xi Wu and Somesh Jha and Ananthram Swami},
      year={2016},
      eprint={1511.04508},
      archivePrefix={arXiv},
      primaryClass={cs.CR}
}

@article{michels2019vulnerability,
  title={On the vulnerability of capsule networks to adversarial attacks},
  author={Michels, Felix and Uelwer, Tobias and Upschulte, Eric and Harmeling, Stefan},
  journal={arXiv preprint arXiv:1906.03612},
  year={2019}
}

@article{kuzlu2022adversarial,
  title={The Adversarial Security Mitigations of mmWave Beamforming Prediction Models using Defensive Distillation and Adversarial Retraining},
  author={Kuzlu, Murat and Catak, Ferhat Ozgur and Cali, Umit and Catak, Evren and Guler, Ozgur},
  journal={arXiv preprint arXiv:2202.08185},
  year={2022}
}

@article{gong2020toward,
  title={Toward smart wireless communications via intelligent reflecting surfaces: A contemporary survey},
  author={Gong, Shimin and Lu, Xiao and Hoang, Dinh Thai and Niyato, Dusit and Shu, Lei and Kim, Dong In and Liang, Ying-Chang},
  journal={IEEE Communications Surveys \& Tutorials},
  volume={22},
  number={4},
  pages={2283--2314},
  year={2020},
  publisher={IEEE}
}

@article{catak2022security,
  title={Security concerns on machine learning solutions for 6G networks in mmWave beam prediction},
  author={Catak, Ferhat Ozgur and Kuzlu, Murat and Catak, Evren and Cali, Umit and Unal, Devrim},
  journal={Physical Communication},
  pages={101626},
  year={2022},
  publisher={Elsevier}
}

@inproceedings{pi2012millimeter,
  title={A millimeter-wave massive MIMO system for next generation mobile broadband},
  author={Pi, Zhouyue and Khan, Farooq},
  booktitle={2012 Conference Record of the Forty Sixth Asilomar Conference on Signals, Systems and Computers (ASILOMAR)},
  pages={693--698},
  year={2012},
  organization={IEEE}
}

@ARTICLE{9259112,
  author={Lin, Yun and Zhao, Haojun and Ma, Xuefei and Tu, Ya and Wang, Meiyu},
  journal={IEEE Transactions on Reliability}, 
  title={Adversarial Attacks in Modulation Recognition With Convolutional Neural Networks}, 
  year={2021},
  volume={70},
  number={1},
  pages={389-401},
  doi={10.1109/TR.2020.3032744}}
  
@article{jiang2021project,
  title={Project Gradient Descent Adversarial Attack against Multisource Remote Sensing Image Scene Classification},
  author={Jiang, Yan and Yin, Guisheng and Yuan, Ye and Da, Qingan},
  journal={Security and Communication Networks},
  volume={2021},
  year={2021},
  publisher={Hindawi}
}

@article{srinivasan2021robustifying,
  title={Robustifying models against adversarial attacks by Langevin dynamics},
  author={Srinivasan, Vignesh and Rohrer, Csaba and Marban, Arturo and M{\"u}ller, Klaus-Robert and Samek, Wojciech and Nakajima, Shinichi},
  journal={Neural Networks},
  year={2021},
  publisher={Elsevier}
}

@article{fostiropoulosrobust,
  title={Robust Defense Against Lp-Norm-Based Attacks by Learning Robust Representations},
  author={Fostiropoulos, Iordanis and Shbita, Basel and Marmarelis, Myrl}
}

@article{alkhateeb2019deepmimo,
  title={Deep{MIMO}: A generic deep learning dataset for millimeter wave and massive {MIMO} applications},
  author={Alkhateeb, Ahmed},
  journal={arXiv preprint arXiv:1902.06435},
  year={2019}
}

@phdthesis{vardhan2021ensemble,
  title={An Ensemble Approach for Explanation-based Adversarial Detection},
  author={Vardhan, Raj},
  year={2021}
}

@misc{RemCom,
  title = {Remcom, {Wireless InSite}},
  howpublished = {\url{http://www.remcom.com/wireless-insite}},
  note = {Accessed: 2021-09-30}
}

@misc{O1_s,
  title = {Deep{MIMO}, {'O1' scenario}},
  howpublished = {\url{https://deepmimo.net/scenarios/o1-scenario/}},
  note = {Accessed: 2021-09-30}
}

}

@misc{I1_s,
  title = {Deep{MIMO}, {'I1' scenario}},
  howpublished = {\url{https://deepmimo.net/scenarios/i1-scenario/}},
  note = {Accessed: 2021-09-30}
}

}

@misc{I3_s,
  title = {Deep{MIMO}, {'I3' scenario}},
  howpublished = {\url{https://deepmimo.net/scenarios/i3-scenario/}},
  note = {Accessed: 2021-09-30}
}

@ARTICLE{2021arXiv210201356B,
       author = {{Bai}, Tao and {Luo}, Jinqi and {Zhao}, Jun and {Wen}, Bihan and {Wang}, Qian},
        title = "{Recent Advances in Adversarial Training for Adversarial Robustness}",
      journal = {arXiv e-prints},
     keywords = {Computer Science - Machine Learning, Computer Science - Artificial Intelligence, Computer Science - Cryptography and Security},
         year = 2021,
        month = feb,
          eid = {arXiv:2102.01356},
        pages = {arXiv:2102.01356},
archivePrefix = {arXiv},
       eprint = {2102.01356},
 primaryClass = {cs.LG},
       adsurl = {https://ui.adsabs.harvard.edu/abs/2021arXiv210201356B},
      adsnote = {Provided by the SAO/NASA Astrophysics Data System}
}

@ARTICLE{2021arXiv210204150F,
       author = {{Faruk Tuna}, Omer and {Ozgur Catak}, Ferhat and {Taner Eskil}, M.},
        title = "{Exploiting epistemic uncertainty of the deep learning models to generate adversarial samples}",
      journal = {arXiv e-prints},
     keywords = {Computer Science - Machine Learning, Computer Science - Artificial Intelligence},
         year = 2021,
        month = feb,
          eid = {arXiv:2102.04150},
        pages = {arXiv:2102.04150},
archivePrefix = {arXiv},
       eprint = {2102.04150},
 primaryClass = {cs.LG},
       adsurl = {https://ui.adsabs.harvard.edu/abs/2021arXiv210204150F},
      adsnote = {Provided by the SAO/NASA Astrophysics Data System}
}
@article{CATAK2017184,
title = {Adaptive filterbank-based multi-carrier waveform design for flexible data rates},
journal = {Computers & Electrical Engineering},
volume = {61},
pages = {184-194},
year = {2017},
issn = {0045-7906},
doi = {https://doi.org/10.1016/j.compeleceng.2016.11.039},
url = {https://www.sciencedirect.com/science/article/pii/S0045790616309648},
author = {Evren Catak and Lutfiye Durak-Ata}
}



@ARTICLE{2014arXiv1412.6572G,
       author = {{Goodfellow}, Ian J. and {Shlens}, Jonathon and {Szegedy}, Christian},
        title = "{Explaining and Harnessing Adversarial Examples}",
      journal = {arXiv e-prints},
     keywords = {Statistics - Machine Learning, Computer Science - Machine Learning},
         year = 2014,
        month = dec,
          eid = {arXiv:1412.6572},
        pages = {arXiv:1412.6572},
archivePrefix = {arXiv},
       eprint = {1412.6572},
 primaryClass = {stat.ML},
       adsurl = {https://ui.adsabs.harvard.edu/abs/2014arXiv1412.6572G},
      adsnote = {Provided by the SAO/NASA Astrophysics Data System}
}

@ARTICLE{2016arXiv160702533K,
       author = {{Kurakin}, Alexey and {Goodfellow}, Ian and {Bengio}, Samy},
        title = "{Adversarial examples in the physical world}",
      journal = {arXiv e-prints},
     keywords = {Computer Science - Computer Vision and Pattern Recognition, Computer Science - Cryptography and Security, Computer Science - Machine Learning, Statistics - Machine Learning},
         year = 2016,
        month = jul,
          eid = {arXiv:1607.02533},
        pages = {arXiv:1607.02533},
archivePrefix = {arXiv},
       eprint = {1607.02533},
 primaryClass = {cs.CV},
       adsurl = {https://ui.adsabs.harvard.edu/abs/2016arXiv160702533K},
      adsnote = {Provided by the SAO/NASA Astrophysics Data System}
}


@ARTICLE{2017arXiv170606083M,
       author = {{Madry}, Aleksander and {Makelov}, Aleksandar and {Schmidt}, Ludwig and {Tsipras}, Dimitris and {Vladu}, Adrian},
        title = "{Towards Deep Learning Models Resistant to Adversarial Attacks}",
      journal = {arXiv e-prints},
     keywords = {Statistics - Machine Learning, Computer Science - Machine Learning, Computer Science - Neural and Evolutionary Computing},
         year = 2017,
        month = jun,
          eid = {arXiv:1706.06083},
        pages = {arXiv:1706.06083},
archivePrefix = {arXiv},
       eprint = {1706.06083},
 primaryClass = {stat.ML},
       adsurl = {https://ui.adsabs.harvard.edu/abs/2017arXiv170606083M},
      adsnote = {Provided by the SAO/NASA Astrophysics Data System}
}

@ARTICLE{2017arXiv171006081D,
       author = {{Dong}, Yinpeng and {Liao}, Fangzhou and {Pang}, Tianyu and {Su}, Hang and {Zhu}, Jun and {Hu}, Xiaolin and {Li}, Jianguo},
        title = "{Boosting Adversarial Attacks with Momentum}",
      journal = {arXiv e-prints},
     keywords = {Computer Science - Machine Learning, Statistics - Machine Learning},
         year = 2017,
        month = oct,
          eid = {arXiv:1710.06081},
        pages = {arXiv:1710.06081},
archivePrefix = {arXiv},
       eprint = {1710.06081},
 primaryClass = {cs.LG},
       adsurl = {https://ui.adsabs.harvard.edu/abs/2017arXiv171006081D},
      adsnote = {Provided by the SAO/NASA Astrophysics Data System}
}
\end{document}